\documentclass[a4paper]{jpconf}
\usepackage{graphicx}
\usepackage{wasysym} 
\usepackage{comment}
\usepackage{hyperref}

\usepackage[dvipsnames]{xcolor}
\begin{document}
\pagenumbering{arabic} 
\setlength{\footskip}{15pt}

\title{An experimental platform for levitated mechanics in space}
\thispagestyle{plain}
\pagestyle{plain}

\author{Jack Homans$^{1}$, Elliot Simcox$^{1}$, Jakub Wardak$^{1}$, Laura da Palma Barbara$^{1}$, Tim M. Fuchs$^{1}$, Rafael Muffato$^{1,2}$, Florence Concepcion$^{3}$, Andrei Dragomir$^{3}$, Christian Vogt$^{4}$, Peter Nisbet-Jones$^{5}$, Christopher Bridges$^{6}$, Hendrik Ulbricht$^{1}$}

\address{1. School of Physics and Astronomy, University of Southampton, SO17 1BJ, Southampton, UK\\
2. Department of Physics, Pontifical Catholic University of Rio de Janeiro, Brazil\\
3. Aquark Technologies, Southampton, UK\\
4. ZARM, University of Bremen, Bremen, Germany\\
5. Twin Paradox Labs, London, UK\\
6. Surrey Space Centre, University of Surrey, Guildford, UK}

\ead{H.Ulbricht@soton.ac.uk}

\begin{abstract}
Conducting levitated mechanical experiments in extreme conditions has long been the aim of researchers, as it allows for the investigation of new fundamental physics phenomena. One of the great frontiers has been sending these experiments into the micro-g environment of space, with multiple proposals calling for such a platform. At the same time, levitated sensors have demonstrated a high sensitivity to external stimuli which will only improve in low-vibrational conditions. This paper describes the development of a technology demonstrator for optical and magnetic trapping experiments in space. Our payload represents the first concrete step towards future missions with aims of probing fundamental physical questions: matter-wave interferometry of nanoparticles to probe the limits of macroscopic quantum mechanics, detection of Dark Matter candidates and gravitational waves to test physics beyond the Standard Model, and accelerometry for Earth-observation.
\end{abstract}

\section{Introduction}
Levitated optomechanical systems and sensors have been the subject of rapid development and discovery as a quantum technology in recent years, with advances being made into both the research of fundamental phenomena and practical applications. A recent comprehensive review by Gonzalez-Ballestero \textit{et al.}~\cite{gonzalez2021levitodynamics} outlines the working principles employed. Levitated mechanical systems are being tailored for research in different fields, such as exploring the limits of quantum mechanics~\cite{millen2020quantum, carlesso2022present}, researching the interplay between quantum mechanics and gravity~\cite{bose2025massive}, probing physics beyond the standard model of particle physics and cosmology such as for detection of Dark Matter and Dark Energy candidates~\cite{moore2021searching, kilian2024dark, amaral2024first, kalia2024ultralight, adelberger2022snowmass}, gravitational waves~\cite{arvanitaki2013detecting, pontin2018levitated} and force sensing \cite{attonewton_force_sensing,electric_force_sensing,2020_optic_lev_accelerometer, fuchs2024measuring}. While terrestrial experiments are capable of strongly isolating test systems from the environment to achieve mechanical quality factors (Q-factors) of $10^{10}$  \cite{highest_Q_factors}, the viability of completely decoupling them from vibrations conducted from the Earth is low. Additionally, the effects of gravity on these systems are non-negligible, especially when matter-wave interferometry is considered where limited free-evolution times and possibly gravitational decoherence hinder advances in research. The next big step is therefore to conduct experiments with levitated systems off-world, as described in a recent vision paper~\cite{belenchia2021test}.\\\\
Dedicated proposals and case studies for nanoparticle interferometry in space have been developed to make a case for testing macroscopic quantum superpositions in space~\cite{gasbarri2021testing}, some as part of the general push for quantum technologies in and for space~\cite{belenchia2022quantum, kaltenbaek2021quantum, bassi2022way}. The first consistent proposal for a space-based levitated optomechanical experiment was made in 2012 in the Macroscopic Quantum Resonators (MAQRO) proposal \cite{MAQRO2012} and has since been updated in 2015 \cite{MAQRO2015update} and 2023 \cite{MAQRO2023} for a dedicated satellite mission at Lagrange point L2. The proposal had a close-up assessment by the European Space Agency (ESA) as part of the concurrent design facility (CDF) study under the name of quantum physics platform (QPPF) in 2018. \\\\
These proposals detail long-term plans for a dedicated deep-space mission to conduct matter-wave interferometry experiments that would explore the high-mass limits of quantum physics. The platform would also enable research into understanding various decoherence mechanisms, including gravitational decoherence~\cite{bassi2017gravitational}. Alongside providing isolation from low-frequency vibrational noise, conducting these experiments in space will provide the following benefits:
\begin{itemize}
  \item Longer coherence and free evolution times are required for observing the quantum properties of a released nanoparticle's spatial evolution. The trapping laser's interaction with the nanoparticle acts as a measurement, causing wave-function collapse such that nanoparticles must be released from the trap to allow them to freely evolve. As the required free evolution time for larger nanoparticles ranges between 100 ms to 100 s \cite{talbot_time_calc}, terrestrial experiments cannot conduct these tests as gravity would accelerate the particles out of the trapping region during this period. In space, this issue is removed, such that experiments are limited solely by the control of the nanoparticle's initial thermal motion \cite{bateman2014near_field_interferometry}.
  \item Increased distance from large gravitational fields whose time dilation affects could cause dephasing between different superposition branches \cite{gravity_decoherence_1,gravity_decoherence_2}.
  \item Exposure to Dark Matter/exotic matter that would be otherwise shielded by the Earth's atmosphere \cite{DM_blocked_terrestial_measurement,space_DM_detector}. Signals from some interstellar Dark Matter candidates such as anisotropic Dark Matter 'wind' may give a directional signal that would be measurable by the levitated test masses.
\end{itemize}
In the context of sensing, force and acceleration sensing by freely falling test masses has a proud history in space under the GRACE~\cite{li2019global}, GOCE~\cite{van2015goce} and LISA Pathfinder missions where satellite-based accelerometers were used for gravimetry with the purpose of satellite geodesy and gravity mapping. LISA Pathfinder delivered one of the most sensitive accelerometers to date~\cite{armano2019lisa}. Levitation-based sensors such as those used in our payload hold promise to deliver sensitive accelerometers operating in the quantum domain at comparably large-mass and benefitting from highly spatially resolved position detection based on quantum optomechanics~\cite{aspelmeyer2014cavity}. \\\\
Here we describe the realisation and testing of a payload to undertake levitated mechanical experiments in space. Through a successful application into the `ESA Payload Masters' competition, a position has been secured with The Exploration Company (TEC) for our experiment to be launched into low Earth orbit in June 2025 for a 3-hour flight to experience approximately 30 minutes of micro-$g$ conditions as part of their `Mission Possible' Nyx flight. The payload has been constructed in collaboration with Twin Paradox Labs, Surrey Space Centre, ZARM, BIAS and Aquark Technologies such that both optical and magnetic traps can be created in a compact volume and the experiments run autonomously, on low power, and without telemetry. We shall discuss the theory underlying the two levitation experiments, the mechanical, optical and electrical design of the payload, the environmental testing it has undergone, and present preliminary data that has been collected from it prior to launch.

\section{Brief Physics of Trapping}
While optical \cite{Jamie_V_Thesis} and magnetic \cite{2022_halbach_1} levitation rely on very different fundamental phenomena, calculating their levitation dynamics is surprisingly similar, so shall be discussed in parallel. Optical and magnetic traps rely respectively on the polarizability $\alpha$ of the silica nanoparticle and effective magnetization $\textbf{M}$ of the diamagnetic graphite sheet being trapped. These properties are defined as:
\begin{equation}
\alpha=4 \pi n^{2} \varepsilon_{0} r^{3} \frac{m^{2}-1}{m^{2}+2},
\label{polarizability_eq}
\end{equation}
\begin{equation}
    \textbf{M}=\frac{1}{\mu_{0}}\mathbf{\chi} \cdot \textbf{B}.
    \label{effective_magnetization}
\end{equation}
In Eq. \ref{polarizability_eq}, $n$ is the nanoparticle's refractive index, $\varepsilon_{0}$ the vacuum permittivity, $r$ the nanoparticle's radius and $m=n/n_m$ the relative refractive index where $n_m$ is the surrounding medium's refractive index. In Eq. \ref{effective_magnetization}, $\mathbf{\chi}$ is the material's magnetic susceptibility tensor, which, in the graphite's body frame, is equal to the tensor's diagonal ($\mathbf{\chi} =\mathrm{diag}\left( \chi_{x},\chi_{y},\chi_{z}\right)$), $\mu_{0}$ is vacuum permeability, and $\textbf{B}$ is the external trapping magnetic field.\\\\
$\alpha$ and $\textbf{M}$ can be used to calculate the trapping forces applied to the nanoparticle and graphite sheet as the optical gradient force $\textbf{F}_{\mathrm{grad}}$ and the magnetic force density $\textbf{f}=\left(\textbf{M}\cdot\nabla\right)\textbf{B}$ acting in the graphite's normal (vertical) plane (see Fig. \ref{magnetic_trap_figure}). 
\begin{equation}
    \textbf{F}_{\mathrm{grad}}=\frac{4 \pi n r^{3}}{c}\left(\frac{m^{2}-1}{m^{2}+2}\right) \nabla I\left(\textbf{r}_{\mathrm{rad}}, \textbf{z}\right),
    \label{FGrad}
\end{equation}
\begin{equation}
    \textbf{f}(\textbf{r})=\frac{1}{2\mu_{0}}\left( \chi_{x}B^{2}_{x}+\chi_{y}B^{2}_{y}+\chi_{z}B^{2}_{z}\right).
\end{equation}
In Eq. \ref{FGrad}, $c$ is the speed of light in a vacuum and $I\left(\mathbf{r}_{\mathrm{rad}}, \mathbf{z}\right)$ is the trapping laser beam's Gaussian intensity profile.\\\\
Using these equations, we can calculate the position $\textbf{r}$ dependent optical and magnetic spring stiffnesses $k$ for each axis $i$ of each trap:
\begin{equation}
        k_{(x, y)}=\frac{4 \alpha(\mathrm{NA})^{4} \pi^{3}}{c \varepsilon_{0} \lambda^{4}} P_{0}, \qquad k_{z}=\frac{2 \alpha(\mathrm{NA})^{6} \pi^{3}}{c \varepsilon_{0} \lambda^{4}} P_{0},
        \label{trap_stiffness}
\end{equation}
\begin{equation}
    k=-\nabla\textbf{f}.
\end{equation}
In Eq. \ref{trap_stiffness}, NA is the numerical aperture of the parabolic mirror that focuses the laser beam to create the optical trapping region, $\lambda$ the trapping laser's wavelength, and $P_{0}$ the trapping laser power.\\\\
Using the values of $k$ and the masses $m$ of the levitated oscillators, the oscillation frequencies can be predicted for each trap using $\omega_{i}= \sqrt{\frac{k_i}{m}}$.

\section{Payload Trap Designs}
\subsection{Optical Trap}
The single-beam optical trap uses a high NA (NA$\sim$0.9) aluminium parabolic mirror (\textit{Wielandts UMPT}), to focus a $\sim79$ mW 1555 nm laser beam, collimated by an output coupler (\textit{Thorlabs PAF2A-18C}), to create the trapping region. Details of the parabola trap can be found in previous works~\cite{Jamie_V_Thesis, vovrosh2017parametric}. A roughened aluminium iris was placed in front of the parabolic mirror to block back-reflections from the mirror's flat mirrored face. The trap is designed to capture $\diameter$300 nm silica nanoparticles (\textit{Bang Laboratories SSD2001}) ejected from a PTFE-coated glass cantilever that will be excited by a piezoelectric actuator (\textit{Piezo Drive SA030310}). The viewport used to provide optical access was anti-reflection coated for 1555 nm (\textit{Torr Scientific Ltd}) to prevent back-reflections off the viewport from obscuring the nanoparticle's signal.

\subsection{Magnetic Trap}
\begin{figure}[h!]
    \centering
    \includegraphics[width=0.9\linewidth]{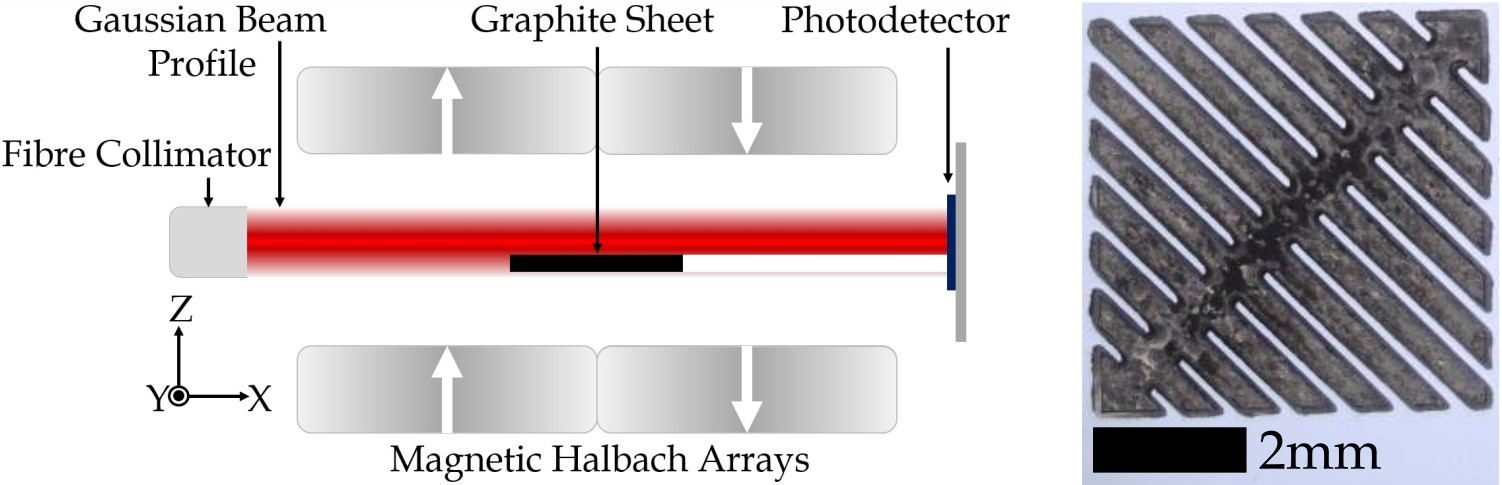}
    \caption{\centering {\bf Magnetic trap layout with optical detection.} \textbf{Left:} An illustration of the dual Halbach array passive magnetic trap with the laser beam offset from the trap's centre. Graphite levitation position is indicated by the black rectangle in the centre. Laser light is blocked by the graphite and detected in transmission on a photodetector. \textbf{Right:} A photograph of pyrolytic graphite sheet laser cut into a fish-bone configuration.}
    \label{magnetic_trap_figure}
\end{figure}

The passive magnetic trap levitates a pyrolytic graphite flake (\textit{K\&J Magnetics Ltd, PG3}) between two 2 $\times$ 2 Halbach arrays of 5 $\times$ 5 $\times$ 2 mm samarium cobalt magnets (\textit{SMMagnetics Sm2Co17-32}) \cite{2022_halbach_1, 2022_halach_2}. In terrestrial experiments, the graphite is held by gravity above a single repelling magnet array. In our setup, the second array will replace gravity and, in micro-$g$ conditions, the graphite shall levitate around the midpoint between the arrays. The arrays are arranged in an attracting configuration as this produces a more stable trapping potential. The viability of this setup has been demonstrated in the temporary micro-$g$ conditions of the ZARM drop-tower.\\\\
The graphite's position is measured using a shadow-detection method, where a $\sim$10 mW $\diameter$1.42 mm 1555 nm laser beam is shone horizontally from a output coupler (\textit{Thorlabs PAF2-A7C}) past the graphite to a large surface area photodetector (\textit{Thorlabs FDG05}), with the laser beam offset from the trap's midpoint to maximize sensitivity. Changes in the graphite's position will therefore proportionally modulate the detected laser power. The trap's configuration (shown in Fig. \ref{magnetic_trap_figure}) provides the greatest sensitivity to the graphite's $z$-displacement, but the graphite's surface roughness and slight misalignment from the horizontal enables detection of its $x$ and $y$ motion.\\\\
The graphite was laser cut into a 4.94 $\times$ 4.95 $\times$ 0.66 ($\pm0.01$) mm square and patterned into a fish-bone configuration (\textit{Laser Micromachining Ltd}). The patterning (shown in Fig. \ref{magnetic_trap_figure}) splits the eddy-currents induced by the graphite's motion relative to the magnets into smaller loops, reducing the forces generated that oppose the graphite's motion and enhancing the Q-factor. The graphite was coated in a thin layer of vacuum compatible epoxy (\textit{Epotek 302-3M}) to increase its strength and durability, and to increase the graphite's mass and therefore its acceleration sensitivity.
\subsection{Vacuum Chamber}
The two traps were placed inside a custom two-chambered CF16 vacuum chamber machined from austenitic stainless steel. The traps were placed in separate chambers to accommodate the different vacuum pressure requirements of each trap ($\sim$0.1 mbar for the optical trap and $\sim10^{-3}$ mbar for the magnetic). The optical trap's chamber has 4 ports, allowing for optical access, electrical feedthrough, piezo positioning and vacuum pumping. The magnetic trap's chamber has 3 ports; two for optical access and one for vacuum pumping.\\\\
Active vacuum pumping and venting to space is not available during the mission, so a non-evaporable getter was coated onto the chambers' internal surfaces (\textit{Aquark Technologies}). This will maintain the chambers' pressures for the 6 month period between payload completion and the mission launch, and throughout the mission. Once the chambers had been evacuated, they were baked-out, both to improve the vacuum and to activate the getter, and then pinched off. Additionally, sapphire viewports (\textit{Lewvac SVP-UV-11.6-16CF}) were used to provide optical access into the chambers, as they are more effective at preventing light gases from leaking into the chambers compared to standard borosilicate viewports.
\begin{figure}
    \centering
    \includegraphics[width=0.9\linewidth]{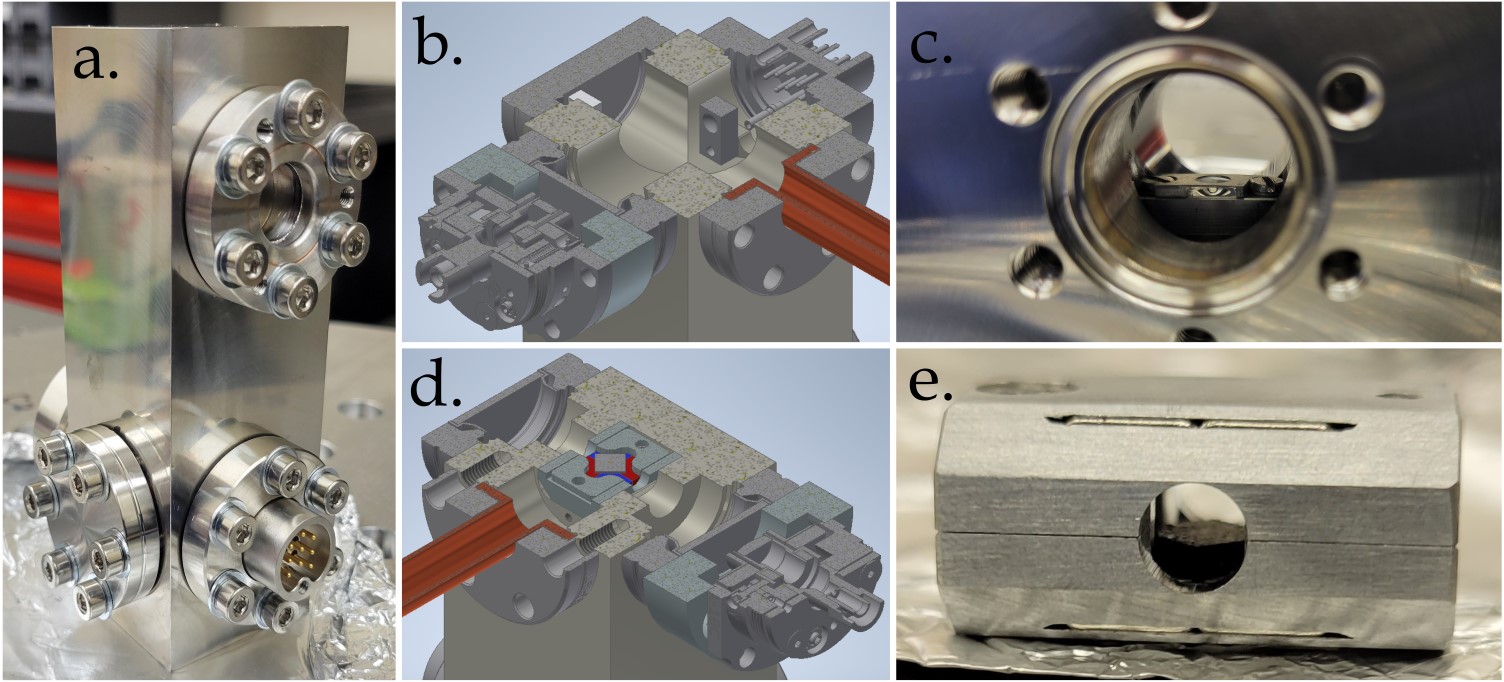}
    \caption{\centering {\bf Details of vacuum chamber with optical and magnetic traps.} \textbf{a.} The custom CF16 vacuum chamber used in the payload. \textbf{b.} A CAD model of the optical trap vacuum chamber, showing the parabolic mirror bolted to the vacuum chamber, opposing the laser collimator. The electrical feedthrough, piezoelectric actuator and copper pinch-off tube are also shown. \textbf{c.} The iris stuck over the top of the parabolic mirror. \textbf{d.} A CAD model of the vacuum chamber used for magnetic trapping. The graphite and one magnet array are shown in their holder, along with the two viewports and the copper pinch-off tube. \textbf{e.} A prototype of the magnetic trap, with the magnets held 4mm apart and the fish-bone pattern, epoxy coated graphite sheet levitating between the two arrays.}
    \label{FM vac chamber}
\end{figure}

\section{Payload Design}
\begin{figure}
    \centering
    \includegraphics[width=0.9\linewidth]{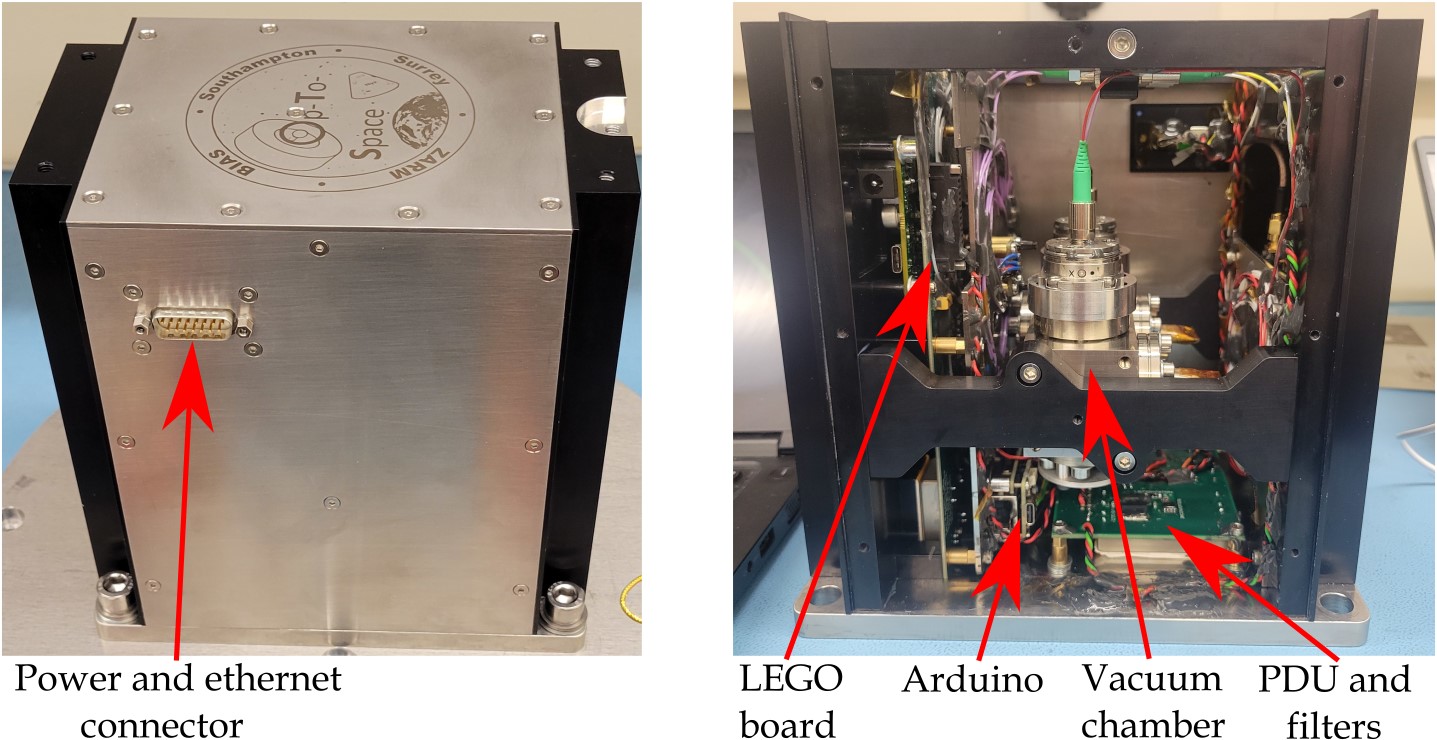}
    \caption{\centering {\bf Photographs of flight model of payload for levitated mechanical experiments in space.} \textbf{Left:} The external form of the payload. \textbf{Right:} The internal layout of components in the payload showing components of the experiments. The system operates fully autonomously and starts experimental protocols upon trigger from spacecraft.}
    \label{payload_images}
\end{figure}

Various parameters were set by The Exploration Company for the design and configuration of the payload (shown in Table \ref{TEC_design_specs}).
\begin{center}
\begin{table}[h]
\centering
\caption{The technical specifications set by The Exploration Company for the payload design.} 
\begin{tabular}{@{}l*{15}{l}}
\br
Parameter&Specification\\
\mr
Volume&20 $\times$ 20 $\times$ 15 cm\\
Mass&$\leq$10 kg\\
Power&1 $\mathrm{W_{av}}$/kg  $\leq10$ $\mathrm{W_{av}}$, 84 $\mathrm{W_{peak}}$, 28 V\\
Data&1 GB main storage, 100 MB backup\\
Thermal conditions&0$\rightarrow$50$^{\circ}$C\\
\br
\label{TEC_design_specs}
\end{tabular}
\end{table}
\end{center}
The payload's weight-dependent power allowance resulted in it being designed to weigh 10 kg to obtain the full 10 W$_{\mathrm{av}}$ power allowance. This meant that, unconventionally, large parts of the payload were made from solid steel and aluminium. If the payload could be exposed to the vacuum of space instead of needing a vacuum chamber and the power-per-weight requirement was removed, the payload's mass could be reduced to $\sim1$ kg.\\\\
One of the payload's end-plates was designed as a heat sync for the laser driver PCB and FPGA (see Section \ref{electronics_section}), while the other was recessed to hold the optical trap's photodetector (see Section \ref{optics_section}). Both end-plates were bolted to the stainless steel baseplate which was designed with M10 through-holes for bolting the payload to the Nyx capsule. The end-plates were also held together at the top of the payload by aluminium stringers, while larger stringers between the two end-plates held the vacuum chamber near the payload's midpoint. Another stringer was added between the top stringers to hold the optical fibre mating sleeves. Stainless steel cover-plates were attached to the stringers and end-plates to enclose the payload. Images of the completed payload can be seen in Fig. \ref{payload_images}. To ensure that all structural components would be secure throughout the mission, all threaded parts used locking threads (\textit{Spiralock}) and a threadlocking agent (\textit{Loctite 242}). Similarly, all internal wires and optical fibres were secured with dots of epoxy (\textit{3M DP190}) at least every 2 cm.

\section{Electronics}
\label{electronics_section}
The laser (\textit{Thorlabs DFB1550P}) used in the experiments is driven by a custom laser driver PCB, produced by a joint team from Twin Paradox Labs and Surrey Space Centre~\cite{chau2022designing,chau2023portable}, that was controlled by an FPGA (\textit{Xilinx Artix 7 A200}). The laser driver enables measurement and control of the laser’s output and thermo-electric controller, and also has capabilities for analogue input and output, PID control and 16-bit data saving. The laser driver and FPGA jointly draws approximately 7 W.\\\\
The 28 V$_{\mathrm{DC}}$ supply from the Nyx capsule is passed through a fifth-order low pass filter to protect both the Nyx capsulse and the payload from voltage fluctuations. The input is then internally converted to a 5 V$_{\mathrm{DC}}$ line and $\pm$12 V$_{\mathrm{DC}}$ rails. The ±12 V$_{\mathrm{DC}}$ is used to power the optical trap's photodetector (\textit{Thorlabs PDBEVAL1}) while the 5 V$_{\mathrm{DC}}$ powers the remaining systems, namely the laser driver PCB and attached FPGA, the optical trap's piezo kicking system and a microcontroller (\textit{Arduino Portenta H7}) used to control internal signals and acquire experimental data. The 5 V$_{\mathrm{DC}}$ also provides a reverse bias to the magnetic trap's photodetector (\textit{Thorlabs FDG05}) to increase the detector's sensitivity.\\\\
The microcontroller has 3 internal analogue-digital converters which are used to collect data from the two photodetectors and a reference accelerometer (\textit{Analogue Devices EVAL-ADXL354BZ}). The data will be collected with 16-bit resolution in 8 MB packets (limited by the microcontroller's RAM) and saved with duplicates on an internal micro-SD card to protect the data from corruption. Before reaching the microcontroller, the photodetectors' signals are passed through band-pass filters, DC-shifted and amplified to maximize the system's sensitivity. The data from the optical and magnetic traps is acquired, respectively, at sampling rates of 500 kSa/s and 10 kSa/s to allow for sufficient bandwidth to measure the predicted oscillation frequencies of each trap. The accelerometer data is only saved alongside the magnetic trap data, as the nanoparticle is too light to effectively measure the same vibrational noise floor.

\section{Optics System and Laser}
\label{optics_section}
A system of optical fibres pass the trapping/monitoring laser around the payload. A 90:10 beam-splitter (\textit{Thorlabs TN1550R2A1}) couples the beam from the laser's output (\textit{Thorlabs DFB1550P}, 100 mW CW at 1555 nm) into the traps, sending 90\% to the optical trap and 10\% to the magnetic trap. The optical trap's beam is then passed through a circulator (\textit{Thorlabs 6015-3-APC}) which circulates the incident light to the optical trap, and couples the returning light into one arm of a balanced photodetector (\textit{Thorlabs PDBEVAL1 with PDB770C}).

\section{Proof of Functionality}
\begin{figure}
    \centering
    \includegraphics[width=\linewidth]{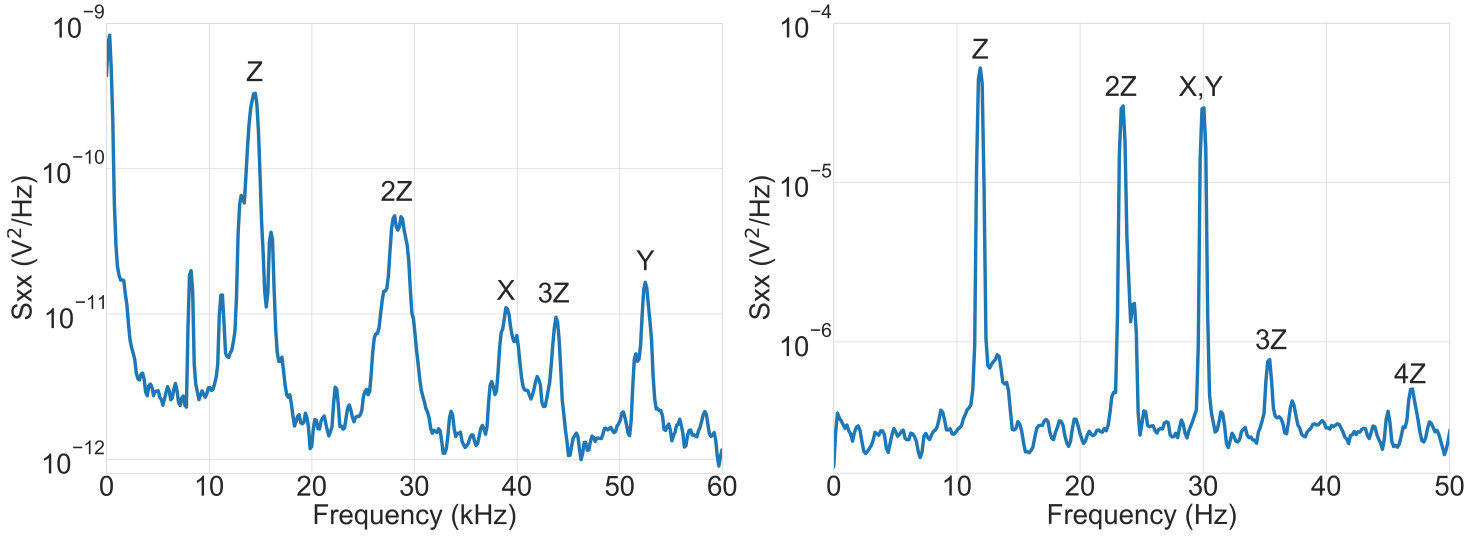}
    \caption{\centering {\bf Demonstration of traps and detection of motion.} Power spectral densities (PSDs) of time traces taken for both traps without active stabilisation and by optical detection. Observed motional frequencies are in good agreement with theoretically modelled ones. \textbf{Left:} The spectrum taken using an identical optical trap and vacuum chamber. A 70 mW trapping laser was used to trap a $\diameter$300 nm silica nanoparticle in 0.46 mbar after it had been loaded into the trap with a nebuliser. The PSD shows translational $x,y,z$-motional degrees and coupling. \textbf{Right:} The PSD of the data taken during the software testing using the magnetic trap in the flight model payload at $\sim10^{-3} $mbar. The peaks in the graph represent the graphite's $x,y,z$-translational modes and coupling.}
    \label{data_figure}
\end{figure}

An identical optical trap was used to trap a $\diameter$300 nm silica nanoparticle. The trap was loaded using a medical nebuliser. Fig. \ref{data_figure} shows a power-spectral density (PSD) recovered from the light returning from the trap. The oscillation frequencies observed match well with the expected range of 10-100 kHz. Additionally, it has been shown that the piezo system that will be used to load the nanoparticles into the trap during autonomous operation is capable of ejecting nanoparticles from its surface. However, to date, we have yet to capture nanoparticles launched using this method. It is believed that the ejected nanoparticles are accelerated by gravity to such a degree that the trap cannot slow them sufficiently. We therefore predict that, in micro-$g$ conditions, the nanoparticles will travel slower towards the trap and be captured.\\\\
While the payload was undergoing software testing (see Section \ref{testing_section}), data was taken with the magnetic trap in the payload using all the systems that will be used throughout the mission. The PSD of the graphite's motion (shown in Fig. \ref{data_figure}) shows that the graphite's multiple distinct oscillation frequencies can be detected and match well with the predicted frequency range of 1-100 Hz.\\\\
The oscillation frequencies of both traps are expected to differ from those presented in Fig. \ref{data_figure} when in micro-$g$ as, in terrestrial experiments, gravity shifts the oscillators away from the trap's centre to a different equilibrium point with different trap spring constants.

\section{Environmental Testing}
\label{testing_section}
To ensure that the payload will operate safely and completely throughout its mission, it has been subjected to multiple rounds of testing. Two versions of the payload were constructed, an engineering model (EM) and a flight model (FM). Electromagnetic (EMC) and thermal testing was conducted on the EM, along with the first round of mechanical testing. 
The payload's design was then modified based on the testing results before the FM was constructed. The FM then underwent the second round of mechanical testing, power characterization testing and software tests to ensure it met the TEC launcher requirements. \\\\
Thermal testing assessed the payload’s responses to changing external thermal environments. The test was conducted at Surrey Space Centre using their Kambie Climatic Chamber (\textit{KK-190 CHULT}). The payload’s functionality was assessed in steps between 0$^{\circ}$C and 40$^{\circ}$C, and the payload was then subjected to thermal stress testing, varying its temperature between -40$^{\circ}$C and +70$^{\circ}$C in seven cycles over 5 days. No effects were observed and the payload retained full functionality.\\\\
The EM underwent three mechanical testing phases. Characterization of mechanical properties involved determining its mass and centre of gravity. Vibrational testing involved conducting resonance searches before and after random vibration tests to determine the payload's fundamental modes and ensure that it will not be adversely affected by the mission's vibrational conditions. Shock testing ensured that the mission's shock environment will not affect the payload's functionality. The FM also had its mechanical properties characterized and underwent acceptance level vibrational tests to assess its build quality and confirm that it was within design tolerances. The EM's vibrational and shock tests and FM's vibrational tests were conducted using Surrey Space Centre's mechanical testing facilities. An \textit{LDS Model V826 shaker head} was used to conduct the vibrational tests in all three of the payload's axes, while a custom shock testing table was used to exert sufficient forces on the EM in each axis. Modifications were made to the payload's design following testing on the EM such that the FM passed all mechanical testing stages.\\\\
EMC testing of the EM was conducted at the BAE Systems Rochester testing facility, and consisted of 4 parts; conducted emission, conducted susceptibility, radiated emission and radiated susceptibility. The testing levels for each of these can be seen in Table \ref{EMC_test_requirements}. The conducted emission and susceptibility tests were passed without issue. Spikes in the payload's emission were detected during the radiated emission test, but were agreed to be acceptable by TEC. The payload was affected by the radiated susceptibility test, but the affected electronic component was identified and replaced with a compatible alternative for the FM.
\begin{center}
\begin{table}[h]
\centering
\caption{EMC testing requirements.} 
\begin{tabular}{p{0.3\linewidth} p{0.65\linewidth}}
\br
Test&Requirements\\
\mr
Conducted Emission&[30;10k] Hz: CE101-4: Curve 2 “28 VOLTS OR BELOW”  [10k;10M] Hz: CE102-1: Basic curve “28V”\\
Conducted Susceptibility&[30;150k] Hz: CS101-1: Curve 2 “28 VOLTS OR BELOW” [10k;200M] Hz: CS114-1: Curve 3 “SPACE”\\
Radiated Emission&TEC custom requirements\\
Radiated Susceptibility&TEC custom requirements\\
\br
\label{EMC_test_requirements}
\end{tabular}
\end{table}
\end{center}
Software testing checked that the payload could interact with the Nyx capsule throughout the mission. The payload was tested in its various operational modes and its response to all possible input signals was assessed. Additionally, the payload's data output was inspected following the test to ensure that all data was collected (see Fig. \ref{data_figure}). The test showed full functionality under all requirements.\\\\
The power characterization test assessed the payload's power usage throughout the mission - specifically the initial current in-rush when switched on, the change in power consumption when the experimental sequence started and terminated, and the power drain when the payload was switched off. Additionally, the payload's thermal output was measured at multiple locations to ensure the Nyx capsule and surrounding payloads would not be affected by the payload. The test showed that the current in-rush was below the required latching current limit and that the average power consumption was 10.94 W - within 10\% of the expected 10 W.

\section{Conclusion and Outlook}
We have described the design and realisation of a space payload for in-orbit levitated mechanical experiments that will be launched in June 2025. The payload contains two types of trap, one optical and one magnetic, which may assist in optimizing success of our first space mission. The payload has passed all tests to specifications for space qualification and has demonstrated full functionality of detection of mechanical modes in both traps. \\\\
Our payload has been designed as a prototype to build on for future missions, with many of the internal systems already capable of more functionality than are currently being used. It is hoped that, in future missions, we will be able to parametrically cool the optically levitated nanoparticle's centre-of-mass motion such that we can increase the spatial coherence in the motional state ready for conducting matter-wave interferometry experiments. Additionally, further missions are in development that aim to use the passive magnetic traps to detect certain Dark Matter candidates.

\section*{Acknowledgments}
We thank James Chalk, Philip Connell, Mark Bampton, Damon Grimsey, Jamie Faux, Oliver Warner, Leigh Allen, Luca Ferrain, Colin Chau, and Robin Elliott for their expert technical support and consultation. We also thank the Space South Central cluster for their support. This project has been made possible by the ESA Payload Master's competition, providing us with a place in The Exploration Company's `Mission Possible' launch. HU acknowledges funding by the EU EIC Pathfinder project QuCoM (101046973). We thank for support the UKRI EPSRC (EP/W007444/1, EP/V000624/1 and EP/X009491/1), the Leverhulme Trust (RPG-2022-57) and the QuantERA II Programme (project LEMAQUME) that has received funding from the European Union’s Horizon 2020 research and innovation programme under Grant Agreement No 101017733. CV acknowledges support from DLR project NaiS (50WM2180).

\section*{References}
\bibliographystyle{ieeetr}
\bibliography{References}

\begin{thebibliography}{10}

\bibitem{gonzalez2021levitodynamics}
C.~Gonzalez-Ballestero, M.~Aspelmeyer, L.~Novotny, R.~Quidant, and O.~Romero-Isart, ``Levitodynamics: Levitation and control of microscopic objects in vacuum,'' {\em Science}, vol.~374, no.~6564, p.~eabg3027, 2021.

\bibitem{millen2020quantum}
J.~Millen and B.~A. Stickler, ``Quantum experiments with microscale particles,'' {\em Contemporary Physics}, vol.~61, no.~3, pp.~155--168, 2020.

\bibitem{carlesso2022present}
M.~Carlesso, S.~Donadi, L.~Ferialdi, M.~Paternostro, H.~Ulbricht, and A.~Bassi, ``Present status and future challenges of non-interferometric tests of collapse models,'' {\em Nature Physics}, vol.~18, no.~3, pp.~243--250, 2022.

\bibitem{bose2025massive}
S.~Bose, I.~Fuentes, A.~A. Geraci, S.~M. Khan, S.~Qvarfort, M.~Rademacher, M.~Rashid, M.~Toro{\v{s}}, H.~Ulbricht, and C.~C. Wanjura, ``Massive quantum systems as interfaces of quantum mechanics and gravity,'' {\em Reviews of Modern Physics}, vol.~97, no.~1, p.~015003, 2025.

\bibitem{moore2021searching}
D.~C. Moore and A.~A. Geraci, ``Searching for new physics using optically levitated sensors,'' {\em Quantum Science and Technology}, vol.~6, no.~1, p.~014008, 2021.

\bibitem{kilian2024dark}
E.~Kilian, M.~Rademacher, J.~M. Gosling, J.~H. Iacoponi, F.~Alder, M.~Toro{\v{s}}, A.~Pontin, C.~Ghag, S.~Bose, T.~S. Monteiro, {\em et~al.}, ``Dark matter searches with levitated sensors,'' {\em arXiv preprint arXiv:2401.17990}, 2024.

\bibitem{amaral2024first}
D.~W. Amaral, D.~G. Uitenbroek, T.~H. Oosterkamp, and C.~D. Tunnell, ``First search for ultralight dark matter using a magnetically levitated particle,'' {\em arXiv preprint arXiv:2409.03814}, 2024.

\bibitem{kalia2024ultralight}
S.~Kalia, D.~Budker, D.~F.~J. Kimball, W.~Ji, Z.~Liu, A.~O. Sushkov, C.~Timberlake, H.~Ulbricht, A.~Vinante, and T.~Wang, ``Ultralight dark matter detection with levitated ferromagnets,'' {\em Physical Review D}, vol.~110, no.~11, p.~115029, 2024.

\bibitem{adelberger2022snowmass}
E.~Adelberger, D.~Budker, R.~Folman, A.~A. Geraci, J.~T. Harke, D.~M. Kaplan, D.~F.~J. Kimball, R.~Lehnert, D.~Moore, G.~W. Morley, {\em et~al.}, ``Snowmass white paper: precision studies of spacetime symmetries and gravitational physics,'' {\em arXiv preprint arXiv:2203.09691}, 2022.

\bibitem{arvanitaki2013detecting}
A.~Arvanitaki and A.~A. Geraci, ``Detecting high-frequency gravitational waves with optically levitated sensors,'' {\em Physical review letters}, vol.~110, no.~7, p.~071105, 2013.

\bibitem{pontin2018levitated}
A.~Pontin, L.~S. Mourounas, A.~A. Geraci, and P.~F. Barker, ``Levitated optomechanics with a fiber fabry--perot interferometer,'' {\em New Journal of Physics}, vol.~20, no.~2, p.~023017, 2018.

\bibitem{attonewton_force_sensing}
G.~Ranjit, D.~P. Atherton, J.~H. Stutz, M.~Cunningham, and A.~A. Geraci, ``Attonewton force detection using microspheres in a dual-beam optical trap in high vacuum,'' {\em Physical Review A}, vol.~91, no.~5, 2015.

\bibitem{electric_force_sensing}
D.~Hempston, J.~Vovrosh, M.~Toroš, G.~Winstone, M.~Rashid, and H.~Ulbricht, ``Force sensing with an optically levitated charged nanoparticle,'' {\em Applied Physics Letters}, vol.~111, no.~13, p.~133111, 2017.

\bibitem{2020_optic_lev_accelerometer}
F.~Monteiro, W.~Li, G.~Afek, C.-l. Li, M.~Mossman, and D.~C. Moore, ``Force and acceleration sensing with optically levitated nanogram masses at microkelvin temperatures,'' {\em Phys. Rev. A}, vol.~101, p.~053835, May 2020.

\bibitem{fuchs2024measuring}
T.~M. Fuchs, D.~G. Uitenbroek, J.~Plugge, N.~van Halteren, J.-P. van Soest, A.~Vinante, H.~Ulbricht, and T.~H. Oosterkamp, ``Measuring gravity with milligram levitated masses,'' {\em Science Advances}, vol.~10, no.~8, p.~eadk2949, 2024.

\bibitem{highest_Q_factors}
L.~Dania, D.~S. Bykov, F.~Goschin, M.~Teller, A.~Kassid, and T.~E. Northup, ``Ultrahigh quality factor of a levitated nanomechanical oscillator,'' {\em Physical Review Letters}, vol.~132, no.~13, p.~133602, 2024.

\bibitem{belenchia2021test}
A.~Belenchia, M.~Carlesso, S.~Donadi, G.~Gasbarri, H.~Ulbricht, A.~Bassi, and M.~Paternostro, ``Test quantum mechanics in space—invest us \$1 billion,'' {\em Nature}, vol.~596, no.~7870, pp.~32--34, 2021.

\bibitem{gasbarri2021testing}
G.~Gasbarri, A.~Belenchia, M.~Carlesso, S.~Donadi, A.~Bassi, R.~Kaltenbaek, M.~Paternostro, and H.~Ulbricht, ``Testing the foundation of quantum physics in space via interferometric and non-interferometric experiments with mesoscopic nanoparticles,'' {\em Communications Physics}, vol.~4, no.~1, p.~155, 2021.

\bibitem{belenchia2022quantum}
A.~Belenchia, M.~Carlesso, {\"O}.~Bayraktar, D.~Dequal, I.~Derkach, G.~Gasbarri, W.~Herr, Y.~L. Li, M.~Rademacher, J.~Sidhu, {\em et~al.}, ``Quantum physics in space,'' {\em Physics Reports}, vol.~951, pp.~1--70, 2022.

\bibitem{kaltenbaek2021quantum}
R.~Kaltenbaek, A.~Acin, L.~Bacsardi, P.~Bianco, P.~Bouyer, E.~Diamanti, C.~Marquardt, Y.~Omar, V.~Pruneri, E.~Rasel, {\em et~al.}, ``Quantum technologies in space,'' {\em Experimental Astronomy}, vol.~51, no.~3, pp.~1677--1694, 2021.

\bibitem{bassi2022way}
A.~Bassi, L.~Cacciapuoti, S.~Capozziello, S.~Dell’Agnello, E.~Diamanti, D.~Giulini, L.~Iess, P.~Jetzer, S.~Joshi, A.~Landragin, {\em et~al.}, ``A way forward for fundamental physics in space,'' {\em npj Microgravity}, vol.~8, no.~1, p.~49, 2022.

\bibitem{MAQRO2012}
R.~Kaltenbaek, G.~Hechenblaikner, N.~Kiesel, O.~Romero-Isart, K.~C. Schwab, U.~Johann, and M.~Aspelmeyer, ``Macroscopic quantum resonators (maqro) testing quantum and gravitational physics with massive mechanical resonators,'' {\em Experimental Astronomy}, vol.~34, pp.~123--164, 2012.

\bibitem{MAQRO2015update}
R.~Kaltenbaek, M.~Aspelmeyer, P.~F. Barker, A.~Bassi, J.~Bateman, K.~Bongs, S.~Bose, C.~Braxmaier, {\v{C}}.~Brukner, B.~Christophe, {\em et~al.}, ``Macroscopic quantum resonators (maqro): 2015 update,'' {\em EPJ Quantum Technology}, vol.~3, pp.~1--47, 2016.

\bibitem{MAQRO2023}
R.~Kaltenbaek, M.~Arndt, M.~Aspelmeyer, P.~F. Barker, A.~Bassi, J.~Bateman, A.~Belenchia, J.~Berg{\'e}, C.~Braxmaier, S.~Bose, {\em et~al.}, ``Research campaign: Macroscopic quantum resonators (maqro),'' {\em Quantum Science and Technology}, vol.~8, no.~1, p.~014006, 2023.

\bibitem{bassi2017gravitational}
A.~Bassi, A.~Gro{\ss}ardt, and H.~Ulbricht, ``Gravitational decoherence,'' {\em Classical and Quantum Gravity}, vol.~34, no.~19, p.~193002, 2017.

\bibitem{talbot_time_calc}
J.~Wardak, T.~Georgescu, G.~Gasbarri, A.~Belenchia, and H.~Ulbricht, ``Nanoparticle interferometer by throw and catch,'' {\em Atoms}, vol.~12, no.~2, p.~7, 2024.

\bibitem{bateman2014near_field_interferometry}
J.~Bateman, S.~Nimmrichter, K.~Hornberger, and H.~Ulbricht, ``Near-field interferometry of a free-falling nanoparticle from a point-like source,'' {\em Nature communications}, vol.~5, no.~1, p.~4788, 2014.

\bibitem{gravity_decoherence_1}
I.~Pikovski, M.~Zych, F.~Costa, and {\v{C}}.~Brukner, ``Universal decoherence due to gravitational time dilation,'' {\em Nature Physics}, vol.~11, no.~8, pp.~668--672, 2015.

\bibitem{gravity_decoherence_2}
M.~Zych, {\em Quantum systems under gravitational time dilation}.
\newblock Springer, 2017.

\bibitem{DM_blocked_terrestial_measurement}
J.~Bateman, I.~McHardy, A.~Merle, T.~R. Morris, and H.~Ulbricht, ``On the existence of low-mass dark matter and its direct detection,'' {\em Scientific reports}, vol.~5, no.~1, p.~8058, 2015.

\bibitem{space_DM_detector}
C.~J. Riedel and I.~Yavin, ``Decoherence as a way to measure extremely soft collisions with dark matter,'' {\em Physical Review D}, vol.~96, no.~2, p.~023007, 2017.

\bibitem{li2019global}
B.~Li, M.~Rodell, S.~Kumar, H.~K. Beaudoing, A.~Getirana, B.~F. Zaitchik, L.~G. de~Goncalves, C.~Cossetin, S.~Bhanja, A.~Mukherjee, {\em et~al.}, ``Global grace data assimilation for groundwater and drought monitoring: Advances and challenges,'' {\em Water Resources Research}, vol.~55, no.~9, pp.~7564--7586, 2019.

\bibitem{van2015goce}
M.~van~der Meijde, R.~Pail, R.~Bingham, and R.~Floberghagen, ``Goce data, models, and applications: A review,'' {\em International Journal of Applied Earth Observation and Geoinformation}, vol.~35, pp.~4--15, 2015.

\bibitem{armano2019lisa}
M.~Armano, H.~Audley, J.~Baird, P.~Binetruy, M.~Born, D.~Bortoluzzi, E.~Castelli, A.~Cavalleri, A.~Cesarini, A.~Cruise, {\em et~al.}, ``Lisa pathfinder performance confirmed in an open-loop configuration: Results from the free-fall actuation mode,'' {\em Physical review letters}, vol.~123, no.~11, p.~111101, 2019.

\bibitem{aspelmeyer2014cavity}
M.~Aspelmeyer, T.~J. Kippenberg, and F.~Marquardt, ``Cavity optomechanics,'' {\em Reviews of Modern Physics}, vol.~86, no.~4, pp.~1391--1452, 2014.

\bibitem{Jamie_V_Thesis}
J.~A. Vovrosh, {\em Parametric feedback cooling and squeezing of optically levitated particles}.
\newblock PhD thesis, University of Southampton, June 2018.

\bibitem{2022_halbach_1}
P.~Romagnoli, R.~Lecamwasam, S.~Tian, J.~Downes, and J.~Twamley, ``Controlling the motional quality factor of a diamagnetically levitated graphite plate,'' {\em arXiv preprint arXiv:2211.08764}, 2022.

\bibitem{vovrosh2017parametric}
J.~Vovrosh, M.~Rashid, D.~Hempston, J.~Bateman, M.~Paternostro, and H.~Ulbricht, ``Parametric feedback cooling of levitated optomechanics in a parabolic mirror trap,'' {\em JOSA B}, vol.~34, no.~7, pp.~1421--1428, 2017.

\bibitem{2022_halach_2}
X.~Chen, S.~K. Ammu, K.~Masania, P.~G. Steeneken, and F.~Alijani, ``Diamagnetic composites for high-q levitating resonators,'' {\em Advanced Science}, vol.~9, no.~32, p.~2203619, 2022.

\bibitem{chau2022designing}
H.~Chau, I.~Boyle, P.~Nisbet-Jones, and C.~Bridges, ``Designing avionics for lasers \& optoelectronics,'' in {\em 4th Symposium on Space Educational Activities}, Universitat Polit{\`e}cnica de Catalunya, 2022.

\bibitem{chau2023portable}
H.~K. Chau, C.~Bridges, and P.~Nisbet-Jones, ``Portable frequency stabilized lasers for quantum technologies using digital techniques,'' {\em IEEE Transactions on Instrumentation and Measurement}, vol.~72, pp.~1--6, 2023.

\end{thebibliography}

\end{document}